\documentclass[reprint,aps,prl,twocolumn ,groupedaddress,nobibnotes]{revtex4-1}
\usepackage{graphicx}

\usepackage{epsfig}
\usepackage{dcolumn}
\usepackage{color}
\usepackage{bm}

\def \L{\Lambda}
\def \g{\gamma}

\def \be{\begin{equation}}
\def \ee{\end{equation}}
\def \ben{\begin{eqnarray}}
\def \een{\end{eqnarray}}

\def \O{\Omega}

\def \p{\partial}
\def \e{\epsilon}

\def \t{\theta}
\def \P{\Phi}
\def \r{\rho}
\def \k{\kappa}
\def \m{\mu}

\def \e{\epsilon}

\begin{document}

\title{{\bf K}-essence Emergent Spacetime as Generalized Vaidya Geometry}

\author{Goutam Manna }
\altaffiliation{goutammanna.pkc@gmail.com}
\affiliation{Department of Physics, Prabhat Kumar College, Contai, Purba Medinipur-721404, India}

\author{Parthasarathi Majumdar}
\altaffiliation{bhpartha@gmail.com}
\affiliation{School of Physical Sciences, Indian Association for the Cultivation of Science, Kolkata-700032, India}

\author{Bivash Majumder}
\altaffiliation{bivashmajumder@gmail.com}
\affiliation{Department of Mathematics, Prabhat Kumar College, Contai, Purba Medinipur-721404, India}

\begin{abstract}

We establish a formal connection between the {\bf K}-essence emergent gravity scenario and generalizations of Vaidya spacetime. Choosing the {\bf K}-essence action to be of the Dirac-Born-Infeld variety, the physical  spacetime to be a general static spherically symmetric black hole and restricting the {\bf K}-essence scalar field to be a function solely of the advanced or the retarded time, we show that the emergent gravity metric resembles closely the generalized Vaidya metrics for null fluid  collapse proposed by Husain. Imposing null energy conditions on the emergent energy-momentum tensor derived from the emergent Einstein equation, restrictions are obtained on the functions characterizing the emergent metric for consistent identification with generalized Vaidya spacetimes. We discuss the possibility of dynamical horizons in the {\bf K-}essence emergent Vaidya spacetime. Admissible explicit black hole background metrics are discussed as examples.
\end{abstract}



\maketitle

\section{Introduction}

The identification of specific {\bf K}-essence emergent gravity models with known spacetimes has been one approach \cite{babi1}-\cite{babi5} to better understand the physical origin of Dark Energy, if the phenomenon of Late Time Acceleration of the Universe is assumed to truly exist, on the basis of recent cosmological observations. An investigation undertaken to identify emergent {\bf K}-essence spacetimes with known gravitational configurations, of  course, goes beyond the original intent of unravelling Dark Energy, since new motivations from analog gravity make such an investigation a useful enterprise. Based on a specific Dirac-Born-Infeld \cite{born1}-\cite{born3} model for the {\bf K}-essence scalar field and specific standard physical black hole spacetimes, one of us (GM) has co-authored papers \cite{gm1,gm2}, establishing conformal invariance of the emergent spacetime with Barriola-Vilenkin and Robinson-Trautman spacetimes. To explore general properties of such a `map' between emergent spacetimes corresponding to physical black hole geometries, and unexpected curved geometries apparently unrelated to those black holes, we focus in this paper on Dirac-Born-Infeld type {\bf K}-essence scalar field models, in the background of general static spherically symmetric black hole spacetimes. Following the standard construction of the emergent composite metric as a function of the physical spacetime metric and the {\bf K}-essence scalar, we obtain a class of emergent metrics which, under certain restrictions on the scalar field, resemble null fluid collapse models as generalizations \cite{husain}, \cite{wang} of the Vaidya spacetime \cite{vai1}-\cite{jg}. We do not address the question as to whether such a relationship between seemingly unconnected spacetime geometries has any deeper significance in either clarifying any aspect of Dark Energy, or indeed providing any insight into gravitational collapse. As alluded to in the abstract, at this point the mapping between the  disparate spacetimes is quite formal. However, detailed computations for the explicit examples of black hole spacetimes may provide some extra insight for the discerning reader.          

The paper is organized as follows : In section 2, we follow ref.s \cite{babi1}-\cite{babi5} to briefly review the construction of the composite emergent metric for a very general {\bf K}-essence scalar field sector in an arbitrary physical spacetime background, not necessarily stationary. Towards the end of this section, we specialize to the precise {\bf K}-essence scalar field action \cite{gm1,gm2} which we actually use in the present work. In the next section, we construct the emergent spacetime metric for a general static spherically symmetric black hole background, with a scalar field restricted to be an arbitrary function of the advanced or retarded Eddington-Finkelstein time, and to be independent of the other variables of the four dimensional spacetime. Such  a choice implies that the composite emergent metric will violate Lorentz invariance. So, at this point it is not clear if the construction will at all lead to anything useful. To investigate this question of utility, we construct the Einstein tensor corresponding to our emergent metric, and compute the components of the {\it emergent energy-momentum tensor} by direct substitution into an emergent Einstein equation. This emergent tensor must obey energy conditions if the emergent geometry is to have any interpretation as a curved spacetime. This requirement is shown to lead to certain restrictions on the functions characterizing the composite metric, i.e., on the function characterizing the background spacetime, as also on the function of the advanced/retarded time characterising the {\bf K}-essence sector. The possibility that the emergent spacetime due to the {\bf K}-essence scalar field can admit dynamical horizons (DH) is discussed, following earlier work on the Vaidya spacetime admitting a dynamical horizon. This is followed in section 4 by results of computations involving several explicit black hole metrics corresponding to the spacetime background. The last section is discussion of our work.   

\section{Review  of {\bf K}-essence and Emergent Gravity}

In this section, we present a short review of the construction of the effective metric for the emergent spacetime corresponding to a general background geometry and a very general {\bf K}-essence scalar field sector. The {\bf K}-essence scalar field $\phi$ minimally coupled to the background spacetime metric $g_{\mu\nu}$ has action \cite{babi1}-\cite{babi5}
\ben
S_{k}[\phi,g_{\mu\nu}]= \int d^{4}x {\sqrt -g} L(X,\phi)
\label{eq:1}
\een
where $X={1\over 2}g^{\mu\nu}\nabla_{\mu}\phi\nabla_{\nu}\phi$.
The energy-momentum tensor is
\ben
T_{\mu\nu}\equiv {2\over \sqrt {-g}}{\delta S_{k}\over \delta g^{\mu\nu}}= L_{X}\nabla_{\mu}\phi\nabla_{\nu}\phi - g_{\mu\nu}L
\label{eq:2}
\een
$L_{\mathrm X}= {dL\over dX},~~ L_{\mathrm XX}= {d^{2}L\over dX^{2}},
~~L_{\mathrm\phi}={dL\over d\phi}$ and  
$\nabla_{\mu}$ is the covariant derivative defined with respect to the gravitational metric $g_{\mu\nu}$.
The scalar field equation of motion is
\ben
-{1\over \sqrt {-g}}{\delta S_{k}\over \delta \phi}= \tilde G^{\mu\nu}\nabla_{\mu}\nabla_{\nu}\phi +2XL_{X\phi}-L_{\phi}=0
\label{eq:3}
\een
where  
\ben
\tilde G^{\mu\nu}\equiv L_{X} g^{\mu\nu} + L_{XX} \nabla ^{\mu}\phi\nabla^{\nu}\phi
\label{eq:4}
\een
and $1+ {2X  L_{XX}\over L_{X}} > 0$.

 We make the conformal transformation
$G^{\mu\nu}\equiv {c_{s}\over L_{x}^{2}}\tilde G^{\mu\nu}$, with
$c_s^{2}(X,\phi)\equiv{(1+2X{L_{XX}\over L_{X}})^{-1}}$. Then the inverse metric of $G^{\mu\nu}$ is   
\ben G_{\mu\nu}={L_{X}\over c_{s}}[g_{\mu\nu}-{c_{s}^{2}}{L_{XX}\over L_{X}}\nabla_{\mu}\phi\nabla_{\nu}\phi] .
\label{eq:5}
\een
A further conformal transformation \cite{gm1,gm2} $\bar G_{\mu\nu}\equiv {c_{s}\over L_{X}}G_{\mu\nu}$ gives
\ben \bar G_{\mu\nu}
={g_{\mu\nu}-{{L_{XX}}\over {L_{X}+2XL_{XX}}}\nabla_{\mu}\phi\nabla_{\nu}\phi}
\label{eq:6}
\een	
Here one must always have $L_{X}\neq 0$ for $c_{s}^{2}$ to be positive definite and only then equations $(1)-(4)$ will be physically meaningful.

It is clear that, for non-trivial spacetime configurations of $\phi$, the emergent metric $G_{\mu\nu}$ is, in general, not conformally equivalent to $g_{\mu\nu}$. So $\phi$ has properties different from canonical scalar fields, with the local causal structure also different from those defined with $g_{\mu\nu}$. Further, if $L$ is not an explicit function of $\phi$ then the equation of motion $(3)$ reduces to;
\ben
-{1\over \sqrt {-g}}{\delta S_{k}\over \delta \phi}
= \bar G^{\mu\nu}\nabla_{\mu}\nabla_{\nu}\phi=0
\label{eq:7}
\een
We shall take the Lagrangian as $L=L(X)=1-V\sqrt{1-2X}$ with $V$ is a constant. 
This is a particular case of the DBI lagrangian \cite{gm1,gm2}, \cite{born1}-\cite{born3}
\ben
L(X,\phi)= 1-V(\phi)\sqrt{1-2X}
\label{eq:8}
\een
for $V(\phi)=V=constant$~~and~~$kinetic ~ energy ~ of~\phi>>V$ i.e.$(\dot\phi)^{2}>>V$. This is typical for the {\bf K}-essence fields where the kinetic energy dominates over the potential energy. Then $c_{s}^{2}(X,\phi)=1-2X$. For scalar fields $\nabla_{\mu}\phi=\partial_{\mu}\phi$. Thus (\ref{eq:6}) becomes
\ben
\bar G_{\mu\nu}= g_{\mu\nu} - \partial _{\mu}\phi\partial_{\nu}\phi
\label{eq:9}
\een
Note the rationale of using two conformal transformations: the first is used to identify the inverse metric $G_{\mu\nu}$, while 
the second realises the mapping onto the   
metric given in $(9)$ for the lagrangian $L(X)=1 -V\sqrt{1-2X}$.

\section{Emergent Spacetime for general spherically symmetric black holes}

The line element corresponding to a general spherically symmetric static (black hole) spacetime is
\ben
ds^{2}=f(r)dt^{2}-f^{-1}(r)dr^{2}-r^{2}d\O^{2}
\label{eq:10}
\een
where $d\O^{2}=d\t^{2}+d\P^{2}$.

We define the tortoise coordinate by the relation $dr^{*}=f^{-1}(r)dr$ and $v=t+\e r^{*}$.
With these definitions, the line element (\ref{eq:10}) reduces to the Eddington-Finkelstein line element
\ben
ds^{2}=f(r)dv^{2}-2\e dvdr-r^{2}d\O^{2}.
\label{eq:11}
\een
When $\e =+1$, the null coordinate $v$ represents the Eddington advanced time (outgoing),  while when $\e =-1$, it represents the Eddington retarded time (incoming). 

Now, from (\ref{eq:9}) the emergent spacetime is described by the line element
\ben
dS^{2}=ds^{2}-\p_{\mu}\phi\p_{\nu}\phi dx^{\mu}dx^{\nu}.
\label{eq:12}
\een
We make the assumptions that the scalar field $\phi(x)=\phi(v)$, so that the emergent spacetime line element is
\ben
dS^{2}=[f(r)-\phi_{v}^{2}]dv^{2}-2\e dvdr-r^{2}d\O^{2}
\label{eq:13}
\een
where $\phi_{v}=\frac{\p \phi}{\p v}$.
Notice that this assumption on $\phi$ actually violates local Lorentz invariance, since in general,
spherical symmetry would only require that $\phi(x)=\phi(v,r)$. The additional assumption that $\phi(v,r)=\phi(v)$ i.e., it is independent of $r$ implies that outside of this particular choice of frame, a spherically symmetric $\phi$ is actually a function of both $v, r$.

We now compare the emergent spacetime (\ref{eq:13}) with the metric \cite{husain,wang} of the generalized Vaidya spacetimes corresponding to gravitational collapse of a null fluid (take $\e=+1$)
\ben
dS_V^2 = \left(1-\frac{2m(v,r)}{r} \right)dv^2 - 2dvdr -r^2 d\O^2
\label{gvai} 
\een
yielding the mass function 
\ben
m(v.r) = \frac12 r\left[1 + \phi_v^2 - f(r) \right].
\label{massf}
\een

One can now compute the emergent Einstein tensor following \cite{husain}, with the notation that subscripts on $m$ and $f$ denote derivatives with respect to those subscripts.
\ben 
{\cal G}^0_0 = {\cal G}^1_1 &=& -\frac{2m_r}{r^2} = \frac{1}{r^2} \left[ rf_r + f -1- \phi_v^2 \right],~ \label{gee00} \\
{\cal G}^1_0 &=& 2\frac{m_v}{r^2} = \frac{2\phi_v \phi_{vv}}{r}, \label{gee10} \\
{\cal G}^2_2 = {\cal G}^3_3 &=& -\frac{m_{rr}}{r} = \frac{1}{2r} \left[2f_r+ rf_{rr} \right]. \label{gee22}  
\een
Recourse to the `emergent' Einstein equation 
\ben
{\cal G}^{\mu}_{\nu} = \k {\cal T}^{\mu}_{\nu}
\een
where, $\k \equiv 8\pi G$ leads to the components ${\cal T}^{\mu}_{ \nu}$, which can be parametrized exactly as in ref.  \cite{husain,wang} in terms of the components $\g, \r$ and $P$ given by
\ben
{\cal T}_{\mu\nu}={\cal T}_{\mu\nu}^{(n)} + {\cal T}_{\mu\nu}^{(m)}=
\left[\begin{array}{cccc}
(\g/2+\r) & \g/2 & 0 & 0	\\
\g/2 & (\g/2-\r) & 0 & 0	\\
0 & 0 & P & 0	\\
0 & 0 & 0 & P
\end{array}\right]\nonumber\\
\label{eq:18}
\een
where ${\cal T}_{\mu\nu}^{(n)}=\g l_{\mu}l_{\nu}$;~ ${\cal T}_{\mu\nu}^{(m)}=(\r+P)(l_{\mu}n_{\nu}+l_{\nu}n_{\mu})+P\bar{G}_{\mu\nu}$
with $l_{\mu}$ and $n_{\mu}$ are two null vectors. Contractions of all indices are performed through the emergent metric $\bar{G}_{\mu \nu}$. The expressions for the three independent components are given by,
\ben
\g &=& \frac{2\phi_v \phi_{vv}}{\k r} \label{gam} \\
\r &=& \frac{1}{\k r^2} \left[ 1 + \phi_v^2 -f -rf_r \right] \label{rho} \\
P &=& \frac{1}{2\k r} \left[2f_r+ rf_{rr} \right]. .\label{pee}
\een
It is obvious that energy conditions imposed on ${\cal T}^{\mu \nu}$ will in turn constrain $f(r)$ and $\phi(v)$ and their derivatives. Thus, 
\ben
\g > 0 &\Rightarrow & \phi_v \phi_{vv} > 0 , \label{gamm} \\
\r > 0 & \Rightarrow & 1 +\phi_v^2 > f + rf_r, \label{rh} \\
P>0 & \Rightarrow & 2f_r + rf_{rr} > 0 . \label{pi}
\een

\subsection{Possible Dynamical Horizon in the Emergent Spacetime}

Having demonstrated that the emergent spacetime is a generalized Vaidya spacetime, obeying the Einstein equation with an emergent energy-momentum tensor  ${\cal T}^{\mu \nu}$ obeying the energy conditions (\ref{gamm})-(\ref{pi}), it stands to reason to enquire, if, just like null hypersurfaces in the ingoing Schwarzschild-Vaidya spacetime discussed in ref. \cite{ash2003} as an example of dynamical horizons, our emergent spacetime also admits dynamical horizons. We rewrite eqn (\ref{eq:13}), given in terms of ingoing Eddington-Finkelstein coordinates $v,r,\theta, \phi$, as       
\ben
dS^{2}=F(v,r) dv^{2}-2 dvdr-r^{2}d\O^{2}
\label{effv}
\een
where, $F(v,r) \equiv f(r) - \phi^2_v(v)$. We follow ref. \cite{ash2002,ash2003,ash2004,hayward,blau,poisson} in the discussion of this subsection.

The spherical symmetry of the emergent spacetime (\ref{effv}) implies that this spacetime can be decomposed locally into the Cartesian product of a two dimensional Lorentzian spacetime spanned by outgoing and ingoing null vectors $l^{\m}$ and $n^{\m}$ respectively, and a spacelike 2-sphere. With our choice of coordinates above, at any fixed point of the Lorentzian spacetime, i.e., on the 2-spheres $v=const.~,~r=const$, the null vectors are chosen as 
\begin{eqnarray}
l &=& \partial_v + \frac12 F(v,r) \partial_r \nonumber \\
n &=& -\partial_r
\end{eqnarray} 
so that, $l^2 =0 =n^2~,~ l^{\m} n_{\m} = 1$. One can compute the expansion of these vectors on the chosen 2-spheres, to obtain,
\begin{eqnarray}
\Theta_l = \frac{F(v,r)}{r}~,~\Theta_n = -\frac{2}{r} \label{expan}
\end{eqnarray}
It is clear from eqn (\ref{expan}) that the hypersurface $F(v,r)=0$ foliated by the chosen 2-spheres, is  a candidate dynamical horizon, since the 2-spheres correspond to {\it marginal} outer trapped surfaces. Note that $F(v,r)=0 \Rightarrow r=r(v)$ implies that the background metric function $f(r(v)) = \phi^2_v(v)$, the square of the $K$-essence scalar field derivative with respect to $v$. Now, the energy condition eqn (\ref{gamm}) implies that $\phi_v^2$ must be a {\it non-decreasing} function of $v$. Further, it is not difficult to show that ${\cal L}_n \Theta_l < 0$ on the chosen 2-spheres. Spherical symmetry in this case implies hypersurface orthogonality (and hence zero rotation) and also vanishing shear. Thus, the $F(v,r)=0$ hypersurface does indeed correspond to a dynamical horizon, as per the definition given in \cite{ash2003}.
The derivatives of  $F(v,r)$ are $F'=\frac{\p F}{\p r}=f'(r)$ and $\dot{F}=\frac{\p F}{\p v}=-2\phi_{v}\phi_{vv}<0$ with the condition (24). There is also the possibility of reaching an equilibrium {\it isolated} horizon, but that would correspond to the $K$-essence scalar field turning completely non-dynamical. 

What is especially interesting in this case, in contrast to the usual Vaidya spacetime, is that the $K$-essence scalar naturally plays the role of the null dust in the generalized Vaidya spacetime. On the dynamical horizon, for every background spacetime characterized by $f(r(v))$, the presence of the $K$-essence scalar field predicts a dynamical horizon    with the scalar field $\phi(v)$ strongly constrained by $\phi^2_v = f(r(v))$. So, if the background is a vacuum or an electrovac black hole spacetime, only for very restricted scalar field solutions can the emergent spacetime admit a dynamical horizon. Thus, since for spherically symmetric black hole backgrounds,  $f(r(v)) > 0$ for the exterior black hole spacetime, there is no restriction as such on $\phi_v^2$ from the above equality between  $\phi_v^2$ and $f(r(v))$. However, as we see below. for specific choices of  $f(r(v))$, more specific restrictions on $\phi_v^2$ appear. These restrictions, while consistent with energy conditions, are specific to dynamical horizons. 

\section{Explicit Examples of Background Spacetime}

\subsection{Schwarzschild Black Hole as background}

Now, we may choose $f(r) = (1-2M/r)$, i.e., the physical spacetime is an exterior {\it Schwarzschild} spacetime. In this
case, the emergent spacetime has the line element
\ben
dS^{2}=[1-\frac{2M}{r}-\phi_{v}^{2}]dv^{2}-2 dvdr-r^{2}d\O^{2}.
\label{eq:14}
\een
From (15), the mass function is
\ben
m(v,r)=M+\frac{r}{2}\phi_v^2.
\label{eq:16}
\een
Therefore, the non-vanishing components of the Einstein tensors are
\ben
\mathcal{G}^{0}_{0}=\mathcal{G}_{1}^{1}=-\frac{2m_r}{r^{2}}=-\frac{\phi_v^2}{r^{2}};~
\mathcal{G}_{0}^{1}=\frac{2m_v}{r^{2}}=\frac{2\phi_v \phi_{vv}}{r};\nonumber\\
\mathcal{G}_{2}^{2}=\mathcal{G}_{3}^{3}=-\frac{m_{rr}}{r}=0~~~~~~~~~~~
\label{eq:17}
\een
where all subscripts designate derivaties as in the last section. For the case of Schwarzschild background (\ref{eq:14}) the values of $\g$, $\r$ and $P$ are
\ben
\g=\frac{2\phi_v \phi_{vv} }{\k r};~
\r=\frac{\phi_v^2}{\k r^{2}}~and~
P=0
\label{eq:20}
\een
which satisfies the weak and strong energy conditions \cite{haw-ellis,wang}
\ben
\g\geq0,~\r\geq0,~P\geq0~ (\g\neq0)
\label{eq:21}
\een
and the dominant energy condition
\ben
\g\geq0,~\r\geq~P\geq 0~(\g\neq 0)
\label{domi}
\een
provided $\phi_{v}\phi_{vv}>0$.
Therefore, energy-momentum tensor becomes
\ben
{\cal T}_{\mu\nu}=
\left[\begin{array}{cccc}
\frac{1}{\k r}(\phi_v \phi_{vv}+\frac{\phi_v^2}{r}) & \frac{\phi_v \phi_{vv}}{\k r} & 0 & 0	\\
\frac{\phi_v \phi_{vv}}{\k r} & \frac{1}{\k r}({\phi_v \phi_{vv}}-\frac{\phi_v^2}{r}) & 0 & 0	\\
0 & 0 & 0 & 0	\\
0 & 0 & 0 & 0
\end{array}\right]\nonumber\\
\label{eq:22}
\een
This type of energy-momentum tensor belongs to type-II class \cite{haw-ellis} which have a double null vector.

The dynamical horizon character for this case with the definition (28) is clear: we get the expansion of the outgoing and ingoing null normals are\ben
\Theta_{l}=\frac{1}{r}(1-\frac{2M}{r}-\phi_{v}^{2});~ \Theta_{n}=-\frac{2}{r}
\een 
Thus, for the dynamical horizon to exist, it is clear that $\phi_v^2 < 1$ in addition to being a non-decreasing function of $v$ as per the energy conditions. This is  a very stringent restriction on what {\bf K}-essence scalar field configurations are permissible for  the generalized Vaidya solution discerned by us, to have a dynamical horizon. 
Also, at the marginally trapped surfaces,
\ben
\mathcal{L}_{n}\Theta_{l}\mid_{F=0}=-\frac{1}{r^{2}}(1-\phi_{v}^{2})<0 ~,~ \phi_v^2 < 1
\een

In this case, $f(r)=1-\frac{r_{S}}{r}$, 
where $r_{S}(=2M)$ is the Schwarzschild radius. For the DH we have
\ben
f(r(v),v)\equiv1-\frac{r_{S}}{r(v)}=\phi_{v}^{2}\nonumber\\
\Rightarrow r(v)=\frac{r_{S}}{1-\phi_{v}^{2}}\simeq r_{S}(1+\phi_{v}^{2}+...)
\label{schdh}
\een
From (39), it is clear that the dynamical horizon corresponding to the emergent metric lies outside the event horizon
of the background black hole. The dynamical horizon location $r(v)$ [FIG.1] is plotted as a function of the K-essence scalar field kinetic energy $\phi_{v}^{2}$ for the admissible range $[0,1]$ and is given in FIG.1.

\begin{figure}[h!]
\centering
\includegraphics[scale=0.6]{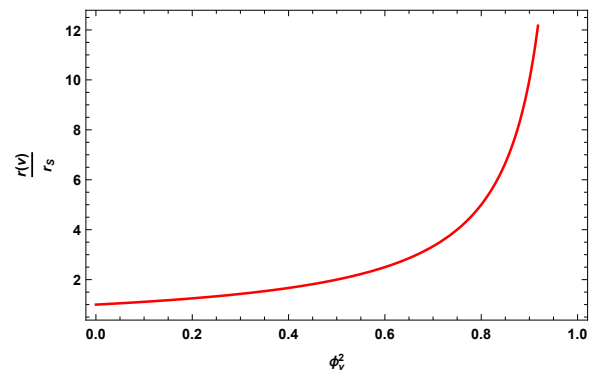}
\caption{(Color online): Plot of the DH location ($r(v)$) (red line) in units of $r_{S}$ vs. $\phi_{v}^{2}$.}
\end{figure}

%\begin{figure}[h!]
%\centering
%\includegraphics[scale=0.6]{figures/sch2.eps}
%\caption{(Color online): Plot the mass function at the horizon $m(r(v),v)$ (blue line) in unit of $r_{S}$ vs. $%\phi_{v}^{2}$.}
%\end{figure}

FIG.1 clearly shows the stretching out of the dynamical horizon of the emergent spacetime beyond the event horizon of the background black hole. This stretching increases with the magnitude of the kinetic energy $\phi_v^2$ of the {\bf K}-essence scalar field, reaching out to asymptopia as the kinetic energy approaches its limiting value  of unity. This delineates the impact of the scalar field on the background spacetime. It also suggests that cosmological effects might begin to arise as this approach to asymptopia of the dynamical horizon occurs. A detailed discussion of these interesting implications is, unfortunately, beyond the scope of the present investigation, where the similarity of the emergent spacetime to a generalized Vaidya spacetime is all we have focused on.   

Next, we consider $\phi_{v}^{2}$ as an explict function of $v$ as 
\ben
\phi_{v}^{2}=e^{-v/v_{0}}\theta(v) \label{phiv}
\een
where $v_{0}$ is a positive constant. This choice of the K-essence scalar field kinetic energy appears to be a reasonable one subject to the restriction on $\phi_v^2$. The idea is to examine the effect of small departures from the background spacetime on the generalized Vaidya spacetime, due to the scalar field interactions. At the dynamical horizon, the mass function (31), in units of the Schwarzschild mass $M$ becomes
\ben
%r(v)=\frac{r_{S}}{1-e^{-v/v_{0}}};\\
\frac{m(r(v),v)}{M}= (1-e^{-v/v_{0}})^{-1}.
\een
A plot the mass function in units of the Schwarzschild black hole mass $M$, as an explicit function of $v$ (with $v_{0}=2$), is given in FIG. 2.

%\begin{figure}[h!]
%\centering
%\includegraphics[scale=0.6]{figures/sch3.eps}
%\caption{(Color online): The DH (red line) in unit of $r_{S}$ against $v$ considering $v_{0}=2$.}
%\end{figure}

\begin{figure}[h!]
\centering
\includegraphics[scale=0.6]{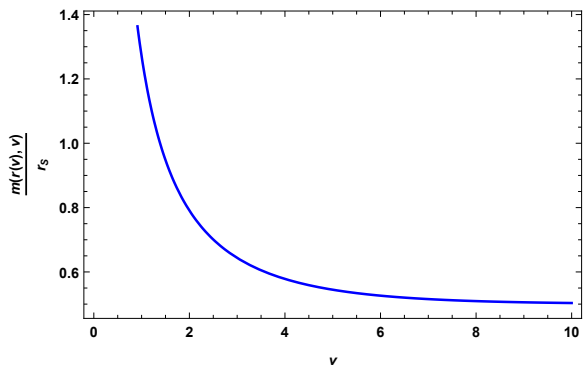}
\caption{(Color online): The mass function (blue line), at the horizon, in unit of $M$ vs $v$ for $v_{0}=2$.}
\end{figure}
The behaviour of the mass function of the generalized Vaidya spacetime is seen to be significantly impacted in the region of large kinetic energy of the scalar field, thus tallying with the behaviour of the location of the dynamical horizon with this kinetic energy. This happens at vanishingly small values of $v$, for the explicit choice depicted in eqn (\ref{phiv}). Other reasonable choices for the explicit functional form of the scalar field ought to lead to the same conclusions.

\subsection{Reissner-Nordstrom Black Hole as background}

Again, we choose $f(r)=1-\frac{2M}{r}+\frac{Q^{2}}{r^{2}}$ i.e., the physical spacetime is {\it Reissner-Nordstrom (RN)} where $Q$ is the charge of the RN black hole. In this case the line element of the emergent spacetime is
\ben
dS^{2}=[1-\frac{2M}{r}+\frac{Q^{2}}{r^{2}}-\phi_{v}^{2}]dv^{2}-2 dvdr-r^{2}d\O^{2}\nonumber\\
\label{eq:23}
\een
and the mass function is
\ben
m(v,r)=M-\frac{Q^{2}}{2r}+\frac{r}{2}\phi_{v}^{2}
\label{eq:24}
\een

For this case the non-vanishing components of Einstein tensors are
\ben
\mathcal{G}^{0}_{0}=\mathcal{G}_{1}^{1}=-\frac{Q^{2}}{r^{4}}-\frac{\phi_{v}^{2}}{r^{2}};~
\mathcal{G}_{0}^{1}=\frac{2\phi_v \phi_{vv}}{r};~
\mathcal{G}_{2}^{2}=\mathcal{G}_{3}^{3}=\frac{Q^{2}}{r^{4}}.\nonumber\\
\label{eq:25}
\een
Using the relation (\ref{eq:24}) we get the values of $\g$, $\r$ and $P$ are
\ben
\g=\frac{2\phi_v \phi_{vv}}{\k r};~
\r=\frac{1}{\k r^{2}}[\frac{Q^{2}}{r^{2}}+\phi_{v}^{2}]~and~
P=\frac{Q^{2}}{\k r^{4}}
\label{eq:26}
\een
which have also satisfied the weak and strong
energy condition (\ref{eq:21})  and dominant energy condition (35) as $\g\geq 0, ~ Q^{2}+r^{2}\phi_{v}^{2}\geq Q^{2}\geq 0$.

In this case, from (28), the expansions are 
\ben
\Theta_{l}=\frac{1}{r}(1-\frac{2M}{r}+\frac{Q^{2}}{r^{2}}-\phi_{v}^{2});~ \Theta_{n}=-\frac{2}{r}
\een 
and 
\ben
\mathcal{L}_{n}\Theta_{l}\mid_{F=0}=-\frac{1}{r^{2}}(1-\frac{Q^{2}}{r^{2}}-\phi_{v}^{2})<0
\een
At the horizon, $F(v,r)=1-\frac{2M}{r}+\frac{Q^{2}}{r^{2}}-\phi_{v}^{2}=0$, we get, exactly as in the Schwarzschild example, the restriction $\phi_v^2 < 1$ for an Reissner-Nordstrom electrovac black hole background. As before, this is the stringent requirement on the {\bf K}-essence scalar field configuration for the emergent RN-Vaidya spacetime to admit a dynamical horizon. In this case, $f(r)=\frac{(r-r_{+})(r-r_{-})}{r^{2}}$, where $r_{\pm}$ are respectively the outer (event) and inner (Cauchy) horizon locations. For the dynamical horizon we have
\ben
&~&\frac{(r(v)-r_{+})(r(v)-r_{-})}{r^{2}(v)}=\phi_{v}^{2}\nonumber\\
&\Rightarrow& r_{\pm}(v)=\frac{1}{2}[(r_{+}+r_{-})\pm \sqrt{(r_{+}-r_{-})^{2}+4r_{+}r_{-}\phi_{v}^{2}}](1+\phi_{v}^{2}+...)\nonumber\\
\label{rndh}
\een
Assuming that the black hole is far from extremality, $r_{+} >> r_{-}$ and also that $\phi_{v}^{2}<1$, we have, the locations of the dynamical horizons as
\ben
r_{\pm}(v)=r_{\pm}(1+\phi_{v}^{2})\pm \frac{r_{+}r_{-}}{r_{+}-r_{-}}\phi_{v}^{2}\pm ...
\label{rndh1}
\een
Once again, we notice that the presence of the {\bf K-}essence scalar field pushes the dynamical horizon of the emergent metric to beyond the event horizon of the background RN black hole. The dynamical horizon location ($r_{+}(v)$) in units of radius of the event horizon of the background black hole $r_{+}$ can be written as
\ben
\frac{r_{+}(v)}{r_{+}}&=&\frac{1+(r_{-}/r_{+})}{2(1-\phi_{v}^{2})}\nonumber\\
&+&\frac{1}{2(1-\phi_{v}^{2})}\sqrt{(1-\frac{r_{-}}{r_{+}})^{2}+4\frac{r_{-}}{r_{+}}\phi_{v}^{2}};~\label{51}\\
\frac{m(r_{+}(v),v)}{r_{+}}&=&\frac{1}{2}(\frac{r_{+}(v)}{r_{+}})(1+\phi_{v}^{2})\nonumber\\&-&\frac{1}{2(r_{+}(v)/r_{+})}(\frac{r_{+}(v)}{r_{+}}-1)(\frac{r_{+}(v)}{r_{+}}-\frac{r_{-}}{r_{+}})\nonumber\\\label{52}
\een

The graphs of the dynamical horizon ($r_{+}(v)$) in units of $r_{+}$ vs $\phi_{v}^{2}$ are given in FIG3, for the non-extremal Reissner-Nordstrom background with two well-separated values of $r_-/r_.$.  

\begin{figure}[h!]
\centering
\includegraphics[scale=0.6]{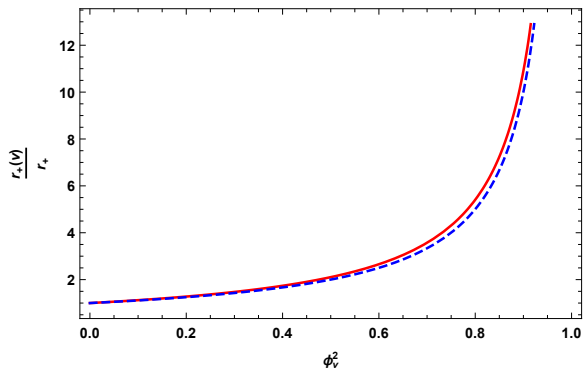}
\caption{(Color online): Plot of the  DH location $r_{+}(v)$ in units of $r_{+}$  vs $\phi_{v}^{2}$, with $\frac{r_{-}}{r_{+}}=0.1$ (red solid line) and $\frac{r_{-}}{r_{+}}=0.001$ (blue dashed line).}
\end{figure}
The graphs show a remarkable similarity for both non-extremal ratios, showing that the inner horizon has little influence on the location of the dynamical horizon, stretched beyond the event horizon of the background black hole spacetime.
%\begin{figure}[h!]
%\centering
%\includegraphics[scale=0.6]{figures/rn3.eps}
%\caption{(Color online): This figure is represent the behaviour of the mass function ($m(r_{+}(v),v)$), at the outer %DH, in unit of $r_{+}$ with the kinetic energy of the {\bf K-}essence scalar field ($\phi_{v}^{2}$) where we consider %$\frac{r_{-}}{r_{+}}=0.01$ (red solid line) and $\frac{r_{-}}{r_{+}}=0.001$ (blue dashed line).}
%\end{figure}

Now, putting the expression of $\phi_{v}^{2}$ (40) in the equation (51), yields the plot of the corresponding mass function at the horizon $m(r_+(v),v)$ in unit of $r_{+}$, as an explicit function of $v$, this is given in FIG4. Once again, the  location of the inner horizon of background appears to have little impact on the mass function.

%\begin{figure}[h!]
%\centering
%\includegraphics[scale=0.6]{figures/rn2.eps}
%\caption{(Color online): Outer DH against $v$ graph in unit of $r_{+}$, considering $\frac{r_{-}}{r_{+}}=0.01$ (red %solid line) and $\frac{r_{-}}{r_{+}}=0.001$ (blue dashed line) and $v_{0}=2$.}
%\end{figure}

\begin{figure}[h!]
\centering
\includegraphics[scale=0.6]{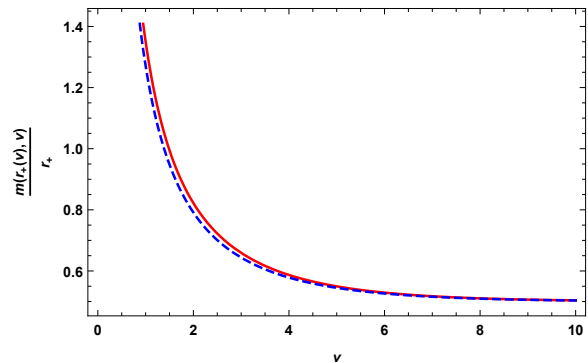}
\caption{(Color online): The mass function, at the DH, vs $v$ graph in unit of $r_{+}$, considering $\frac{r_{-}}{r_{+}}=0.01$ (red solid line) and $\frac{r_{-}}{r_{+}}=0.001$ (blue dashed line) and $v_{0}=2$.}
\end{figure}

\vspace{3in}

{\bf Extremal Reissner-Nordstrom black hole background:}
In this case, $r_{+}= r_{-}\equiv r_{0}$, so that $f (r) = (r - r_{0} )^{2} /r^{2}$. It is easy to see that, even though the background metric
does not have a {\it bifurcate} horizon, the DH of the corresponding emergent spacetime appears to have a bifurcate structure, given by the locations 
\ben
\frac{(r(v)-r_{0})^{2}}{r^{2}(v)}=\phi_{v}^{2}\nonumber\\
\Rightarrow r_{\pm}(v)=r_{0}(1\mp \phi_{v})^{-1}
\een
The outer horizon, once again has been pushed out compared to the event horizon of the background black hole. The plot of $r_{+}(v)$ in units of $r_{0}$ vs $\phi_{v}$ is given in FIG5.

\begin{figure}[h!]
\centering
\includegraphics[scale=0.6]{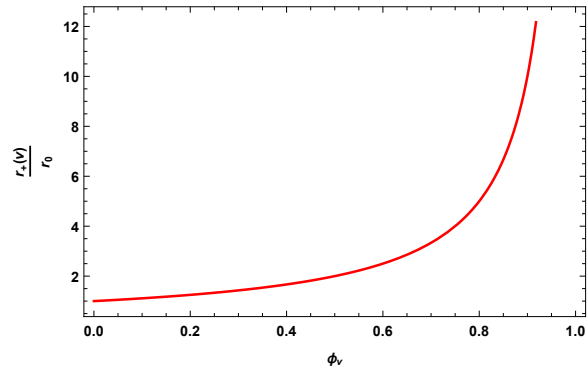}
\caption{(Color online): The extremal DH location ($r_{+}(v)$) in units of $r_{0}$ vs. $\phi_{v}$ graph.}
\end{figure}
%\begin{figure}[h!]
%\centering
%\includegraphics[scale=0.6]{figures/xrn2.eps}
%\caption{(Color online): The extremal outer DH ($r_{+}(v)$) in unit of $r_{0}$ vs. $v$ (using equation (40)) graph considering $v_{0}=2$.}
%\end{figure}

The mass function in this extremal case can be expressed in terms of the explicit form of $\phi_v^2$, given by eqn (\ref{phiv})
\begin{eqnarray}
\frac{m(r_{+}(v),v)}{r_{0}} &=& \frac{1}{2(1-e^{-v/2v_{0}})}
\end{eqnarray}

The plot of the mass function (53) in unit of $r_{0}$, at the extremal outer DH, against $v$ is given in FIG. 6.

\begin{figure}[h!]
\centering
\includegraphics[scale=0.6]{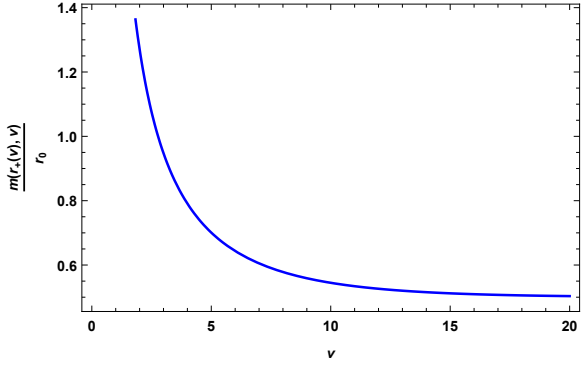}
\caption{(Color online): The mass function at the outer DH location ($r_{+}(v)$) in unit of $r_{0}$ vs. $v$ graph considering $v_{0}=2$.}
\end{figure}

\subsection{de-Sitter Reissner-Nordstrom black hole as background}

If we consider $f(r)=(1-\frac{2M}{r}+\frac{Q^2}{r^2}-\frac{\L}{3}r^2)$ where $\L$ is the cosmological constant i.e., the physical spacetime is Reissner-Nordstrom-de Sitter (RNdS) ($\L>0$) then the emergent spacetime is
\ben
dS^{2}=[1-\frac{2M}{r}+\frac{Q^2}{r^2}-\frac{\L}{3}r^2-\phi_{v}^{2}]dv^{2}-2 dvdr-r^{2}d\O^{2}\nonumber\\
\label{eq:28}
\een
and the mass function becomes
\ben
m(v,r)=M-\frac{Q^{2}}{2r}+\frac{\L}{6}r^{3}+\frac{r}{2}\phi_{v}^{2}
\label{eq:29}
\een
and non-zero components of Einstein tensors are
\ben
\mathcal{G}^{0}_{0}=\mathcal{G}_{1}^{1}=-[\frac{Q^{2}}{r^{4}}+\L+\frac{\phi_{v}^{2}}{r^{2}}];~
\mathcal{G}_{0}^{1}=\frac{2\phi_v \phi_{vv}}{r};\nonumber\\
\mathcal{G}_{2}^{2}=\mathcal{G}_{3}^{3}=\frac{Q^{2}}{r^{4}}-\L.
\label{eq:30}
\een
In this case the values of $\g$, $\r$ and $P$ are
\ben
\g=\frac{2\phi_v \phi_{vv}}{\k r};~
\r=\frac{1}{\k r^{2}}[\frac{Q^{2}}{r^{2}}+\L r^{2}+\phi_{v}^{2}]~and~\nonumber\\
P=\frac{1}{\k}[\frac{Q^{2}}{r^{4}}-\L]
\label{eq:31}
\een
which have to satisfy the energy conditions (\ref{eq:21}) provided $\frac{Q^{2}}{r^{4}}>\L$ and dominant energy condition (35) as $\g\geq 0, ~ Q^{2}+\L r^{4}+r^{2}\phi_{v}^{2}\geq Q^{2}-\L r^{4}\geq 0$.

In this case, from (28), the expansions are 
\ben
\Theta_{l}=\frac{1}{r}(1-\frac{2M}{r}+\frac{Q^{2}}{r^{2}}-\frac{\Lambda}{3}r^{2}-\phi_{v}^{2});~ \Theta_{n}=-\frac{2}{r}
\een 
and 
\ben
\mathcal{L}_{n}\Theta_{l}\mid_{F=0}=-\frac{1}{r^{2}}(1-\frac{Q^{2}}{r^{2}}-\Lambda r-\phi_{v}^{2})<0
\een

We have not given any plots in this case, because it is difficult to extract the location of the dynamical horizon from the quartic equation which comes out of the computations.

\section{Discussion}

In the cosmologial context of dark energy, if the kinetic energy of the {\bf K-}essence scalar field $\phi_{v}^{2}$ is a real and positive constant $K$, then the following scenarios arise: (a) for a Schwarzschild background, the emergent spacetime analogoues to the Barriola-Vilenkin type spacetime \cite{gm1} for a particular solution of {\bf K-}essence scalar field. In this case, the global monopole charge is replaced by the constant kinetic energy $(K)$ of the scalar field with $K \in \{0,1 \}$. (b) For Reissner-Nordstrom and de-Sitter Reissner-Nordstrom backgrounds, the emergent spacetime is similar to the Robinson-Trautman (RT) type spacetime \cite{gm2,gm3}, if $K$ is constrained to be unity for a particular solution of {\bf K-}essence scalar field. The event horizons are changed in the above three cases in the pres-
ence of constant kinetic energy of the K-essence scalar
field.

We hasten to add, however, that the link discerned by us between a specific {\bf K}-essence emergent gravity model and generalized Vaidya models characterizing gravitational collapse of null fluids with a large class of mass functions is interesting from a purely gravitational theory standpoint, rather than in the somewhat mystifying cosmological context of dark energy whose very existence may now be in some doubt \cite{subir}, on the basis  of the latest analysis of data from the Planck consortium \cite{planck}. However, the connection found in this paper may have uses in another context, namely the context of analogue gravity models where gravitational phenomena difficult to observe in real spacetime may still be scrutinized in terrestrial laboratories. 

The most striking of analogue models is the class of acoustic black hole analogues \cite{visser} where inviscid, barotropic fluids with specific flow parameters, perturbed by linear acoustic perturbations, have been shown to reproduce experimentally accessible laboratory analogues of Hawking radiation \cite{unruh}, superradiance \cite{bm2003} and inertial frame dragging with Lense-Thirring precession \cite{cgm2017}. Some of these analogue phenomena have been claimed to have been observed \cite{stein}, \cite{wein}. However, the context of such observation is not always within the theoretical premise of vanishing viscosity set down in the incipient works. In fact, nonzero, albeit small viscosity has been shown \cite{visser} to lead to violation of Lorentz invariance, and hence the impossibility of extracting a cogent acoustic metric in these more realistic situations. Even so, it has been shown that superradiance may still be viable if the inclusion of effects due to it can be treated perturbatively in the viscosity coefficient \cite{oindrila}. In this case, it is most meaningful to investigate whether an approximate analogue acoustic metric can indeed be constructed. This is perhaps an arena where the formal connection found in the text above can be of substantial utility, since the configurations of perturbations in such fluids have similarities to the scalar fields in the {\bf K}-essence scenarios discussed above. We hope to report on this interesting relation in the near future.

{\bf Acknowledgement:} 
The authors would like to thank the referees for illuminating suggestions to improve the manuscript.

\end{document}